\documentclass[printer]{aa} 
\usepackage{graphicx}
\usepackage{txfonts}
\usepackage[round]{natbib}
\bibpunct{(}{)}{;}{a}{}{,} 
\usepackage{epsfig}
\usepackage{longtable}

\begin{document}
   \title{Exploring the circumstellar environment of the young
     eruptive\\ star V2492 Cyg\thanks{This work is based on
       observations made with the {\it Herschel} Space Observatory and
       with the \emph{Spitzer} Space Telescope. \emph{Herschel} is an
       ESA space observatory with science instruments provided by
       European-led Principal Investigator consortia and with
       important participation from NASA. \emph{Spitzer} is operated
       by the Jet Propulsion Laboratory, California Institute of
       Technology under a contract with NASA.}}

   \author{\'A.~K\'osp\'al\inst{1}\thanks{ESA fellow}
           \and
           P.~\'Abrah\'am\inst{2}
           \and
           J.~A.~Acosta-Pulido\inst{3,4}
           \and
           M.~J.~Ar\'evalo Morales\inst{3,4}
           \and
           Z.~Balog\inst{5}
           \and
           M.~I.~Carnerero\inst{3,4}
           \and
           E.~Szegedi-Elek\inst{2}
           \and
           A.~Farkas\inst{2}
           \and
           Th.~Henning\inst{5}
           \and
           J.~Kelemen\inst{2}
           \and
           T.~Kov\'acs\inst{2}
           \and
           M.~Kun\inst{2}
           \and
           G.~Marton\inst{2}
           \and
           Sz.~M\'esz\'aros\inst{3,4}
           \and
           A.~Mo\'or\inst{2}
           \and
           A.~P\'al\inst{2,6}
           \and
           K.~S\'arneczky\inst{2,7}
           \and
           R.~Szak\'ats\inst{2}
           \and
           N.~Szalai\inst{2}
           \and
           A.~Szing\inst{2}
           \and
           I.~T\'oth\inst{2}
           \and
           N.~J.~Turner\inst{8}
           \and
           K.~Vida\inst{2}
           }

   \institute{Research and Scientific Support Department, European
     Space Agency (ESA, ESTEC, SRE-SA), PO Box 299, 2200 AG, Noordwijk,
     The Netherlands\\
     \email{akospal@rssd.esa.int}
     \and
     Konkoly Observatory, Research Centre for Astronomy and
     Earth Sciences, Hungarian Academy of Sciences, 
     PO Box 67,\\ 1525 Budapest, Hungary
     \and
     Instituto de Astrof\'\i{}sica de Canarias, 38200 La
     Laguna, Tenerife, Spain
     \and
     Departamento de Astrof\'\i{}sica, Universidad de La Laguna,
     38205 La Laguna, Tenerife, Spain
     \and
     Max-Planck-Institut f\"ur Astronomie, K\"onigstuhl 17,
     69117 Heidelberg, Germany
     \and
     Department of Astronomy, Lor\'and E\"otv\"os University,
     P\'azm\'any P. st. 1/A, Budapest, 1117, Hungary
     \and
     ELTE Gothard -- Lend\"ulet Research Group, 9700 Szombathely,
     Hungary
     \and
     Jet Propulsion Laboratory, California Institute of
     Technology, Pasadena, CA 91109, USA
     }

   \date{Received 13 October 2012 / Accepted 3 January 2013 }

 
  \abstract
  {V2492\,Cyg is a young eruptive star that went into outburst in
    2010. The near-infrared color changes observed since the outburst
    peak suggest that the source belongs to a newly defined sub-class
    of young eruptive stars, where time-dependent accretion and
    variable line-of-sight extinction play a combined role in the flux
    changes.}
  {In order to learn about the origin of the light variations and to
    explore the circumstellar and interstellar environment of
    V2492\,Cyg, we monitored the source at ten different wavelengths,
    between 0.55$\,\mu$m and 2.2$\,\mu$m from the ground and between
    3.6$\,\mu$m and 160$\,\mu$m from space.}
  {We analyze the light curves and study the color--color diagrams via
    comparison with the standard reddening path. We examine the
    structure of the molecular cloud hosting V2492\,Cyg by computing
    temperature and optical depth maps from the far-infrared data.}
  {We find that the shapes of the light curves at different
    wavelengths are strictly self-similar and that the observed
    variability is related to a single physical process, most likely
    variable extinction. We suggest that the central source is
    episodically occulted by a dense dust cloud in the inner disk, and
    based on the invariability of the far-infrared fluxes, we propose
    that it is a long-lived rather than a transient structure. In some
    respects, V2492\,Cyg can be regarded as a young, embedded analog
    of UX Orionis-type stars.}
  {The example of V2492\,Cyg demonstrates that the light variations of
    young eruptive stars are not exclusively related to changing
    accretion. The variability provided information on an azimuthally
    asymmetric structural element in the inner disk. Such an
    asymmetric density distribution in the terrestrial zone may also
    have consequences for the initial conditions of planet formation.}


   \titlerunning{The circumstellar environment of the young eruptive
     star V2492\,Cyg}
   \authorrunning{\'A.~K\'osp\'al et al.}

   \maketitle

\section{Introduction}

Young eruptive stars, defined by their rapid and powerful brightening
in optical and infrared light, form a subclass of Sun-like pre-main
sequence objects. The eruptions of the well-studied classical
FU\,Orionis- and EX\,Lupi-type stars are believed to be due to
enhanced accretion from the circumstellar disk onto the star
\citep{hk96}. According to the current picture, the inward spiraling
material piles up close to the inner edge of the accretion disk and
falls onto the stellar surface as a result of a gravitational and
thermal instability \citep{zhu2009}. This runaway accretion model,
however, may not explain all eruptive events. Recent studies of some
young eruptive stars suggest that the flux changes of these systems
are due to the combination of two effects of comparable amplitude. One
effect is an intrinsic brightening related to the appearance of a new,
accretion-fueled hot component in the system, while the other is a
dust-clearing event that reduces the extinction along the line of
sight (see, e.g., V1647\,Ori in \citealt{reipurth2004} and
\citealt{acosta2007}, or PV\,Cep in \citealt{kun2011}). The
simultaneity of the accretion and extinction changes suggests that
they might be physically linked, and it is the changing accretion
luminosity that causes changes in the inner disk structure. Such
rearrangements of the inner disk have a potentially high importance
for the evolution of the terrestrial zone of circumstellar disks.

V2492\,Cyg (also known as VSX\,J205126.1+440523, IRAS\,20496+4354, and
PTF\,10nvg) is a recently discovered, young eruptive star that may
show another example of extinction-related flux changes.  The source
is located in the Pelican Nebula, at a distance of 550\,pc
\citep{straizys1989,bally2003}. It brightened by 1.8\,mag in
unfiltered light between 2009 December and 2010 June, but Digitized
Sky Survey plates show that it had been several magnitudes fainter in
quiescence \citep{itagaki2010, munari2010}. Light curves presented in
\citet{covey2011} and \citet[][hereafter Paper\,I]{kospal2011}
indicated that the star reached maximum brightness at the end of 2010
August, then showed a fast decline, making its light curve very
similar to that of the outburst of EX\,Lup in 2008 (\'Abrah\'am et
al.~2009). However, further monitoring revealed that the outburst is
not over yet: V2492\,Cyg started to brighten again at the end of
2010. Changes in the near-infrared colors suggested that the initial
brightening leading to the peak in 2010 August was probably due to the
combination of increasing accretion and decreasing
extinction. However, subsequent flux variations mostly followed the
standard interstellar reddening path, indicating that after the peak,
changing extinction was largely responsible for the observed
variability \citep[Paper\,I,][]{aspin2011}. This makes V2492\,Cyg
remarkably similar to V1647\,Ori and PV\,Cep, but with an
unprecedentedly large amplitude in the $A_V$ changes. Further
photometric and spectroscopic monitoring of the object is presented in
\citet{hillenbrand2012}.

In this paper, we present new optical, near-infrared, mid-infrared,
and far-infrared monitoring observations, as well as a near-infrared
spectrum of V2492\,Cyg. In Sect.~\ref{sec:obs} we summarize the
observations, in Sect.~\ref{sec:res} we present light curves,
color--color and color--magnitude diagrams, multi-epoch spectral
energy distributions (SEDs), and the near-infrared spectrum, as well
as our far-infrared results on the source and its surroundings. In
Sect.~\ref{sec:dis} we discuss the physical origin of the flux
variations, and in Sect.~\ref{sec:con} we summarize our conclusions
about the nature and circumstellar environment of V2492\,Cyg.

\section{Observations and data reduction}
\label{sec:obs}

\begin{figure*}
\centering \includegraphics[height=\textwidth,angle=90]{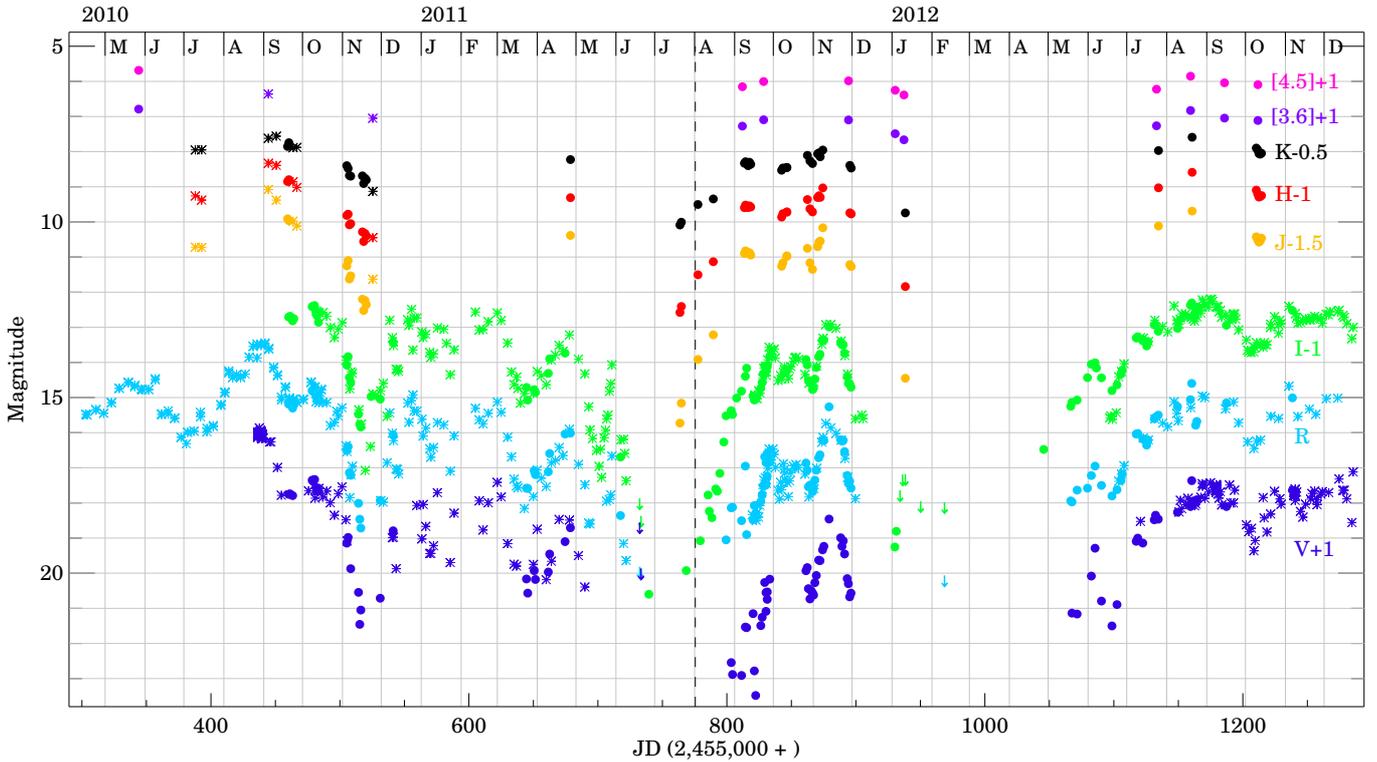}
\caption{Light curves of V2492\,Cyg. Filled dots are from Paper\,I and
  this work, asterisks are from \citet{covey2011}, \citet{aspin2011},
  and from the AAVSO database (http://www.aavso.org). For clarity, the
  $V$, $I$, $J$, $H$, $K_{\rm S}$, [3.6], and [4.5] light curves are
  shifted along the y axis. Downward arrows indicate 3$\sigma$ upper
  limits (this work). Tick marks on the top indicate the first day of
  each month. Vertical dashed line marks the epoch when our WHT/LIRIS
  spectum was taken.}
\label{fig:light}
\end{figure*}

We obtained optical and near-infrared images with $VRIJHK_{\rm S}$
filters between 2010 September 19 and 2012 October 12 using four
telescopes: the Schmidt and RCC telescopes of the Konkoly Observatory
(Hungary), as well as the IAC-80 and Telescopio \emph{Carlos
  S\'anchez} (TCS) telescopes of the Teide Observatory in the Canary
Islands (Spain). Technical details of the telescopes and their
instrumentation are described in Paper\,I. Reduction of the images and
aperture photometry were performed in the same way as in Paper\,I. The
resulting magnitudes for the period between 2010 September 19 and 2011
January 2 are presented in Paper\,I, while the rest are listed in
Table~\ref{tab:phot} and plotted in Fig.~\ref{fig:light}.

We observed V2492\,Cyg using the \emph{Spitzer} Space Telescope in the
post-helium phase at eight epochs between 2011 September 8 and 2012
September 16 (PID: 80165, PI: P.~\'Abrah\'am). We used the IRAC
instrument at 3.6 and 4.5$\,\mu$m in subarray mode. The data reduction
procedures were the same as described in detail in
\citet{kun2011}. Fluxes were color-corrected by convolving the IRAC
filter profiles with the observed SED in an iterative way. The results
of the photometry are listed in Table~\ref{tab:spitzerherschel}.

We conducted a four-epoch far-infrared monitoring of V2492\,Cyg with
the \emph{Herschel} Space Observatory between 2011 October 29 and 2012
January 11 (PID: DDT\_akospal\_2, PI: \'A.~K\'osp\'al). We used the
Photodetector Array Camera and Spectrometer (PACS) instrument
\citep{poglitsch2010} in mini-scan map mode with medium scan speed
(20$''$ s$^{-1}$) at 70$\,\mu$m and 160$\,\mu$m. Two scan maps with
orientations of 70$^{\circ}$ and 110$^{\circ}$ with respect to the
detector array were performed at each epoch. We used eight scan legs
in each map with a 4$''$ separation between them, and the length of
the scan legs was 3$\farcm$0. The dates were scheduled so that the
\emph{Spitzer} and the \emph{Herschel} observations were executed
within two days of each other. Additionally, we obtained 100$\,\mu$m
and 160$\,\mu$m photometry with PACS, as well as 250, 350, and
500$\,\mu$m images with the Spectral and Photometric Imaging Receiver
\citep[SPIRE,][]{griffin2010} at one single epoch. The PACS maps were
taken in the same observation setup as described above. The SPIRE
observations were performed in the small map mode, the repetition
factor was set to 1.

Due to the extended emission surrounding V2492\,Cyg, special care was
taken when extracting photometry. In order to remove the background
from the PACS images, we used the ``boloSource()'' routine (Vavrek et
al., in prep.). This algorithm has been developed to subtract point-
and compact sources from the diffuse background of large-scale
galactic maps observed with the PACS and SPIRE photometers. Unlike
other algorithms working on the projected Level\,2 maps, it takes
Level\,2 masks around pre-identified sources, which are then
back-projected onto the Level\,1 detector pixel timeline. These
timeline segments are interpolated at spatial frequencies adjusted to
the source size (for a point source this is variable with scan
speed). The precise size and position of the Level\,1 mask are further
optimized on the timeline to ensure that the interpolation happens
over the most significant sequence of frames for each individual pixel
and for any specific source. The interpolator first simulates the
high-frequency part of the flux-calibrated signal timeline (that
includes correlated noise, instrument 1/f noise, and extended
background sky emission) with a power spectrum similar to that
measured on a single scan-leg. In a second step, this synthetic noise
component is added to a baseline estimate (low-frequency component) at
the position of masked frames, resulting in an interpolated signal in
the detector timeline without the presence of masked sources. The
background-removed timeline can then be created by subtracting the
interpolated timeline from the original timeline. The projection of
such a timeline provides an image containing the sources only, without
any contribution from the sky background.

To make reliable maps of the extended emission seen in the Herschel
PACS and SPIRE images, we used \emph{scanamorphos}
\citep{roussel2012}, a software developed especially for the purpose
of computing maps from scan observations by taking into account the
redundancy of the data. We found that the maps produced by
scanamorphos contain less striping and result in smoother images than
those produced by the \emph{Herschel} Interactive Processing
Environment \citep[HIPE, ][]{ott2010}.

We used the background-subtracted PACS images to obtain aperture
photometry for V2492\,Cyg and another point source in the field,
HH\,570, by averaging flux values calculated for aperture radii
between 5$''$ and 10$''$ at 70$\,\mu$m and between 14$''$ and 19$''$
at 160$\,\mu$m (due to the source being marginally resolved, see also
Sect.~\ref{sec:extent}). At 70$\,\mu$m, we note that a faint source is
detected at a distance of 9$\farcs$9 from V2492\,Cyg (see also
Sect.~\ref{sec:broader}), but the contamination to the photometry is
less than 3\%. We applied aperture corrections given in
\citet{muller2011}. In order to achieve the highest photometric
accuracy and cancel out any remaining systematic calibration
artifacts, we determined the flux of V2492\,Cyg relative to
HH\,570. For HH\,570, the average of the four epochs provided a flux
value of 4.87$\pm$0.18\,Jy at 70$\,\mu$m and 9.60$\pm$0.52\,Jy at
160$\,\mu$m. We estimated uncertainties by taking into account the
formal error of the aperture photometry, the uncertainty of the
absolute flux calibration (3\%, 3\%, and 5\% at 70, 100, and
160$\,\mu$m respectively, see \citealt{muller2011}), the uncertainty
of the photometry of HH\,570, and the uncertainty in the background
subtraction. Data analysis of SPIRE maps was performed in HIPE
v.~9.0.0, working on the pipeline-processed products. Following the
recommendation of the SPIRE Data Reduction Guide, we used the
sourceExtractorTimeline task (formerly known as Timeline Fitter) to
obtain photometry for V2492\,Cyg. This algorithm works on the Level\,1
baseline-subtracted timeline data rather than on the final maps.  The
task subtracts a 2-degree polynomial baseline from the individual
scans and fits a Gaussian to the source. V2492\,Cyg is situated in a
region with a relatively bright and inhomogeneous background
emission. Although we did what we could to eliminate the background by
doing photometry on the background-subtracted PACS maps and fitting
Gaussians in the timeline data for SPIRE, uncertainty in the
background estimation dominates the final photometric error. We
color-corrected the Herschel fluxes by convolving the PACS and SPIRE
filter profiles with the observed SED in an iterative way. The final
Herschel fluxes and uncertainties for V2492\,Cyg are listed in
Table~\ref{tab:spitzerherschel}.

We observed V2492\,Cyg with the LIRIS instrument installed on the
4.2\,m \emph{William Herschel} Telescope (WHT) at the Observatorio del
Roque de Los Muchachos (Spain). The description of the instrument can
be found in \citet{acosta2007}. On 2011 July 21/22 $JHK_{\rm S}$
images were taken, while on 2011 August 1/2 long-slit intermediate
resolution spectra in the $ZJ$ and $HK$ bands were obtained. The
images were taken in a 5-point dither pattern, with 5\,s of exposure
time per dither position. The spectra were taken in an ABBA nodding
pattern with a total exposure time of 600\,s and 336\,s in the $ZJ$
and $HK$ bands, respectively. We used the 0$\farcs$75 slit width,
which yielded a spectral resolution of R=550--700 in the
0.9--2.4$\,\mu$m range. We observed HIP\,103694, a G2-type star, as
telluric calibrator. The data reduction of both the images and the
spectra were done in the same way as in \citet{acosta2007}. The
spectra were flux calibrated using $JHK_{\rm S}$ photometry taken with
TCS on 2011 August 3/4. The typical signal-to-noise ratio of the
spectra is 5, 25, and 50 in the $J$, $H$, and $K_{\rm S}$ bands,
respectively. The photometry is included in Table~\ref{tab:phot},
while the spectra are plotted in Fig.~\ref{fig:liris}.

V2492\,Cyg was observed by the Wide-field Infrared Survey Explorer
(WISE; \citealt{wright2010}) on 2010 May 27-28. This measurement was
taken during the cryogenic phase and is presented in the WISE All-Sky
Database. For the W1, W2, and W3 filters, 12-16\% of the pixels were
saturated. While in most cases the profile-fitting algorithm of WISE
producing flux estimates from the unsaturated pixels works very well,
there is a pronounced flux-overestimation bias for the W2 filter in
case of sources brighter than W2 = 6.5\,mag. Using the curve in Fig.~8
of Sect.~6.3 of the Explanatory Supplement to the WISE All-Sky Data
Release Products \citep{cutri2012}, we corrected for this bias. Then,
we converted the magnitudes to fluxes and color-corrected them by
convolving the WISE filter profiles with the observed SED in an
iterative way. In the errors, we added in quadrature 2.4\%, 2.8\%,
4.5\%, and 5.7\% as the uncertainty of the absolute calibration in the
W1, W2, W3, and W4 bands, respectively (Sect. 4.4 of the Explanatory
Supplement). The resulting fluxes are $F_{3.353\,\mu\rm
  m}=1.36\pm0.07$\,Jy, $F_{4.603\,\mu\rm m}=2.40\pm0.10$\,Jy,
$F_{11.561\,\mu\rm m}=4.90\pm0.22$\,Jy, and $F_{22.088\,\mu\rm
  m}=7.67\pm0.45$\,Jy.

\section{Results and analysis}
\label{sec:res}

\subsection{Light curves}
\label{sec:light}

Using our measurements from Tables~\ref{tab:phot} and
\ref{tab:spitzerherschel}, along with data from the literature
(including Paper\,I) and from the AAVSO database, we plotted light
curves of V2492\,Cyg taken with eight different filters between 0.55
and 4.5$\,\mu$m (Fig.~\ref{fig:light}). As the $R$-band light curve
shows, the peak of the outburst occurred in 2010 around the end of
August or beginning of September. The highest $J$, $H$, $K_{\rm S}$,
and 3.4--3.8$\,\mu$m fluxes were also detected in 2010 September. The
well-sampled optical light curves show that the object has been
alternating between fading and brightening with amplitudes of several
magnitudes, and no general fading trend can be seen. An exceptionally
deep dip occurred around 2011 June--July, when the $I$-band flux
dropped by an unprecedented 7\,mag. In 2011 November, the source
reached an $R$-band brightness only $\approx$2\,mag below the outburst
peak in 2010 August. In 2011 mid-December, V2492\,Cyg started a rapid
fading, decreasing by at least 6\,mag in $I$-band, then remained
invisible for our telescopes for several months. Photometry from 2012
indicates brightening and another peak in the light curves at the end
of 2012 August, when the source became almost as bright as at
maximum. According to both our latest measurements from 2012 October
and optical observations from the AAVSO data base until the end of
2012, V2492\,Cyg is currently fading again. Analyzing optical
observations of \citet{covey2011} taken between 2010 May and November,
\citet{aspin2011} found multiple brightness peaks with an $\approx$100
day period. Our data in Fig.~\ref{fig:light} exhibit several different
variability timescales, the longest one suggested by the two local
peaks in 2011 November and 2012 August. On shorter timescales, between
2011 August and December, the source seems to show
$\approx$20-day-long quasi-periodic cycles. Recently,
\citet{hillenbrand2012} presented well-sampled $JHKs$ light curves and
claimed to have found a 221-day period.

\begin{figure}
\centering \includegraphics[width=\columnwidth,angle=0]{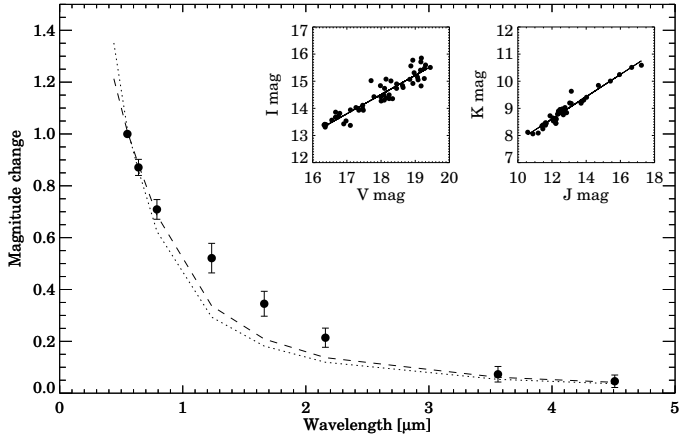}
\caption{Magnitude changes of V2492\,Cyg relative to the
  $V$-band. Interstellar reddening for $R_V$=3.1 is overplotted with
  dotted line and for $R_V$=5.0 with dashed line
  \citep{cardelli1989}. The small insets show examples for scatter
  plots from which the relative magnitude changes were determined as
  the slope of the fitted lines.}
\label{fig:scatter}
\end{figure}

The near- and mid-infrared light curves show a remarkable similarity
in shape to the optical ones, albeit with a lower amplitude. By
correlating the different light curves, we can exclude any time lags
larger than a few days between them. In order to quantify the
variability amplitudes as a function of wavelength, we produced $I$
vs.~$V$, $R$ vs.~$I$, $J$ vs.~$I$, $H$ vs.~$J$, $K_{\rm S}$ vs.~$J$,
[3.6] vs.~$K_{\rm S}$, and [4.5] vs.~[3.6] scatter plots, always using
only those data points for each plot that were obtained on the same
night. Two example plots can be seen as small insets in
Fig.~\ref{fig:scatter}. The scatter plots could be well fitted with
linear relationships, no matter whether the source was fading or
brightening. The slopes of the lines, plotted in
Fig.~\ref{fig:scatter}, determine the variability amplitudes relative
to the $V$-band (0.871$\pm$0.031, 0.709$\pm$0.038, 0.521$\pm$0.057,
0.354$\pm$0.048, 0.214$\pm$0.037, 0.073$\pm$0.030, 0.046$\pm$0.024 for
$R$, $I$, $J$, $H$, $K_{\rm S}$, [3.6], and [4.5], respectively). The
amplitudes monotonically decrease towards longer wavelengths, more
steeply at optical wavelengths than in the infrared. At 4.5$\,\mu$m,
the flux changes are only 5\% of those in the V-band, but are still
very well detected.

\subsection{Herschel monitoring}
\label{sec:herschelmonitoring}

In order to check whether photometric variability can be seen at
far-infrared wavelengths, we compared the Herschel fluxes of
V2492\,Cyg measured at four different epochs
(Table~\ref{tab:spitzerherschel}). Our data reveal that the
peak-to-peak variability of the source is less than 0.6\,Jy (4\%) at
70$\,\mu$m, and less than 0.8\,Jy (5\%) at 160$\,\mu$m. These limits
are comparable to or less than the uncertainty of the individual flux
measurements, we thus conclude that V2492\,Cyg did not show
significant flux changes in the far-infrared between 2011 October and
2012 January.

\subsection{Color--color and color--magnitude diagrams}
\label{sec:cc}

\begin{figure}
\centering \includegraphics[width=\columnwidth,angle=0]{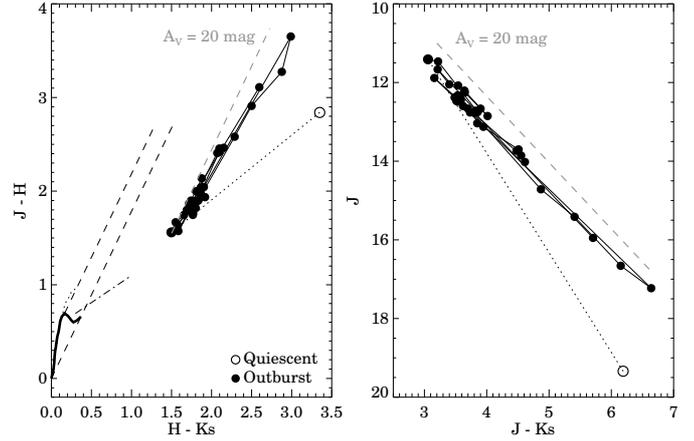}
\caption{Color--color and color--magnitude diagrams of V2492\,Cyg. The
  open circle represents quiescent (2006) data from Paper\,I, while
  filled dots mark observations taken in 2010--2012 from Paper\,I and
  from this work. The main sequence is marked by a thick solid line,
  the giant branch with a dotted line \citep{koornneef1983}, the
  reddening path for $R_V$=3.1 with dashed lines \citep{cardelli1989},
  and the T\,Tauri locus with a dash-dotted line \citep{meyer1997}.}
\label{fig:tcd}
\end{figure}

Based on photometry from 2010, we presented a $J-H$ vs. $H-K_{\rm S}$
color--color diagram in Paper\,I. We found that the source was very red
before the outburst, was bluest at the peak of the outburst, and
started fading approximately along the reddening path \citep[see
  also][]{aspin2011}. Our new data from 2011--2012 confirm that all
color changes happened along a well-defined line, with a slope of
1.40$\pm$0.04, that was close to, but somewhat different from, the
value of 1.78 representative for the standard interstellar reddening
(Fig.~\ref{fig:tcd}, left). Figure~\ref{fig:tcd}, right, suggests that
the relative changes in the $J$ and $K_{\rm S}$ bands are fully
consistent with the extinction law; thus the difference in the
color--color diagram is related to larger-than-expected changes in the
$H$ band with respect to the reddening law. If the near-infrared
magnitude changes were caused by changing extinction, then the
difference in $A_V$ between the peak of the outburst and our faintest
observation in 2011 July would be about 20\,mag. We note that
\citet{hillenbrand2012} observed the source at an even fainter state,
indicating that the full range of extinction changes can be even
higher. Remarkably, the faintest colors from 2010--2012 are still
significantly different from the quiescent colors measured in
2006. The $J$ vs.~$J-K_{\rm S}$ color--magnitude diagram
(Fig.~\ref{fig:tcd}, right) shows that not only the colors, but also
the magnitude changes follow the reddening path, suggesting the same
$\Delta A_V \approx$ 20\,mag. This is different from what was observed
for V1647\,Ori, where the color changes were consistent with
reddening, but the magnitudes required additional color-independent
flux changes as well \citep{acosta2007}.

At optical wavelengths, the distribution of data points in the $V-R$
vs.~ $R-I$ and the $I$ vs.~$V-I$ diagrams are also consistent with
changing extinction. At the \emph{Spitzer} wavelengths, the [3.6]
vs.~[3.6]--[4.5] diagram also follows very precisely the reddening
path. However, the extinction changes indicated by the optical and
mid-infrared data are significantly smaller than the corresponding
value estimated at near-infrared wavelengths: only $\approx$13\,mag
instead of $\approx$20\,mag between the peak and 2011 July. This
discrepancy is reflected in Fig.~\ref{fig:scatter}, where the relative
magnitude changes at the $JHK_{\rm S}$ bands are higher than the
extinction curves fitted to the optical and the \emph{Spitzer} data
points.

\subsection{Spectral energy distributions}
\label{sec:sed}

\begin{figure}
\centering \includegraphics[width=\columnwidth,angle=0]{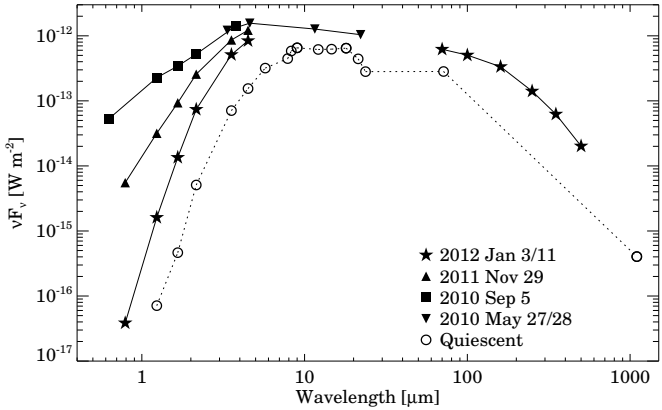}
\caption{Multi-epoch SEDs of V2492\,Cyg. Open circles represent
  quiescent data from Paper\,I and references therein, except for the
  1100$\,\mu$m point, which is from \citet{aspin2011}. Although these
  points are not contemporaneous, we connected them with a dotted line
  to guide the eye. Filled symbols show outburst observations at four
  different epochs. Data from 2010 May are from the WISE catalog, data
  from 2010 September are from \citet{covey2011} and
  \citet{aspin2011}, and data from 2011-2012 are from this work.}
\label{fig:sed}
\end{figure}

The quiescent SED (open circles in Fig.~\ref{fig:sed}) resembles that
of a moderately reddened embedded Class\,I source, with a bolometric
luminosity of 14\,L$_{\odot}$ and bolometric temperature of 280\,K
\citep[Paper\,I;][]{covey2011,aspin2011}. For the outburst period, we
plotted several SEDs, snapshots of the system in different brightness
states. Although we do not have a complete contemporaneous SED
measured at peak brightness in 2010 August, we can attempt to
construct a reasonable approximation of it by simply taking the
``upper envelope'' of the data points plotted in
Fig.~\ref{fig:sed}. For the shortest wavelengths, we take the 2010
September data, which were observed close to the peak. For the
mid-infrared part, we take the WISE data from 2010 May because the
$R$-band light curve shows that the source was in a very similar
brightness state in 2010 May and in 2010 September. For the
long-wavelength part, we take the Herschel data from 2012 January
because our Herschel monitoring (Sect.~\ref{sec:herschelmonitoring})
showed that the source has a constant flux at $\geq$70$\,\mu$m,
independent of the actual optical brightness. Integrating this upper
envelope, the resulting bolometric luminosity and temperature are
43\,L$_{\odot}$ and 570\,K, respectively.

Comparing the quiescent SED with the upper envelope SED (i.e., the SED
representative of the peak brightness), it is evident that an extra
emission component appeared in the system that caused brightening in
the whole optical-to-far-infrared regime, even at 70$\,\mu$m. Such an
outburst component was observed in a number of eruptive objects, such
as OO\,Ser \citep{kospal2007}, V1647\,Ori \citep{muzerolle2005}, and
EX\,Lup \citep{juhasz2012}. Assuming that this component is located in
the center of the system, it is probably reddened by both
circumstellar and interstellar extinction. We took the difference of
the quiescent and the 2010 September SED between 1.25 and 3.8$\,\mu$m
and dereddened it by different values in the 6--12\,mag range
\citep{covey2011}. We found that, irrespective of the applied
dereddening, the outburst component seems to have a temperature
distribution rather than a single temperature blackbody (as opposed to
EX\,Lup or HBC\,722, \citealt{juhasz2012, kospal2011}).

The multi-epoch SEDs plotted for the outburst period deviate from each
other in the optical and near-infrared regime, but converge to the
same flux level in the mid-infrared. This behavior suggests that the
temperature and the emission of the disk, and thus its heating via
irradiation by the central source or the release of accretion energy,
was nearly constant during the outburst. However, the intrinsically
constant light curve of the source is intermittently modulated by
various amounts of fading. As we discussed in Sect.~\ref{sec:cc}, this
modulation is probably related to changing extinction along the line
of sight.

\subsection{Spectroscopy}

\begin{figure}
\centering \includegraphics[width=\columnwidth,angle=0]{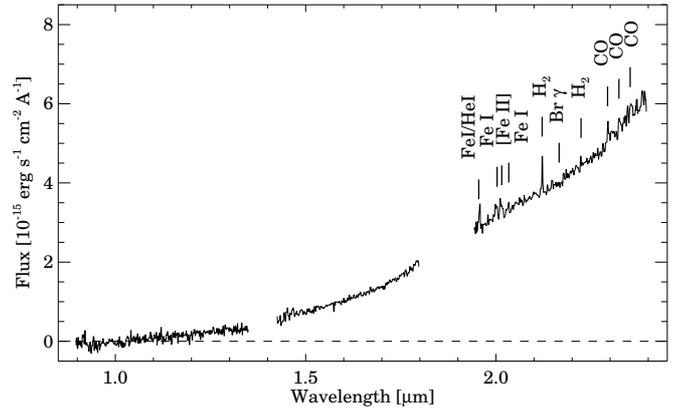}
\caption{Flux-calibrated $ZJ$ and $HK$ spectra of V2492\,Cyg taken on
  2011 August 1 with LIRIS on the WHT.}
\label{fig:liris}
\end{figure}

The $ZJ$ and $HK$ spectra in Fig.~\ref{fig:liris} were taken on 2011
August 1, when V2492\,Cyg was rapidly brightening, but was still close
to its faintest detected state (see the dashed line in
Fig.~\ref{fig:light}). The source was practically undetected below
1.1$\,\mu$m, then the spectrum steeply rose towards longer
wavelengths. The $J$ and $H$ bands show a smooth continuum, no lines
are visible within the measurement uncertainties. The $K$ band,
however, shows some emission features, notably the CO bandhead, two
H$_2$ lines, and some marginally detected metallic lines. No atomic
hydrogen lines are present. $JHK$ spectra were also obtained by
\citet{covey2011} in 2010 July 14 and 18, as well as by
\citet{aspin2011} in 2010 October 2-3 and 2010 November 26-27. Our
spectrum differs in many ways from the earlier spectra. While the
Brackett and Paschen series were conspicuous in 2010, they are absent
in our spectrum. We estimate a 1$\sigma$ upper limit of
4$\times$10$^{-15}$\,erg\,s$^{-1}$\,cm$^{-2}$ for the flux of the
Br$\,\gamma$ line. The CO bandhead emission also became fainter, but
is still detected in our spectrum. The equivalent width (EW) and the
flux integrated between 2.293$\,\mu$m and 2.300$\,\mu$m are
$-$4.5$\AA$ and 2.2$\times$10$^{-14}$\,erg\,s$^{-1}$\,cm$^{-2}$,
respectively. This is a factor of 2--6 decrease in EW and a factor of
3--19 decrease in flux. We note that we did not apply any reddening
correction to the line fluxes.

Interestingly, the H$_2$ lines in the $K$ band exhibit a different
behavior. The lines at 2.12$\,\mu$m and 2.22$\,\mu$m have EWs of
$-$6.7$\,\AA$ and $-$1.4$\,\AA$ and fluxes of
2.5$\times$10$^{-14}$\,erg\,s$^{-1}$\,cm$^{-2}$ and
6.1$\times$10$^{-15}$\,erg\,s$^{-1}$\,cm$^{-2}$, respectively. This
means that the fluxes of the H$_2$ lines became 1.1--2.5 times
stronger than in 2010, while the ratio of the 2.12$\,\mu$m line to the
2.22$\,\mu$m line stays roughly constant over the different
epochs. The ratio, which is equal to 4.1 for our spectrum, is close to
the theoretical value of 4.4 for C shocks computed by
\citet{smith1995}. This suggests that, as opposed to the H\,I lines
and the CO bandhead, the H$_2$ lines are not originating from close to
the central star, but farther from it, probably from a jet.

\citet{hillenbrand2012} present several spectra taken in 2011. Three
of them were observed close in time (2011 June 26, July 15, and August
17) to our 2011 August 1 spectrum. The absolute flux level of our
spectrum is most similar to their 2011 June 26 spectrum, while the
features resemble those seen in their 2011 August 17 spectrum.

\subsection{Spatial extent of V2492\,Cyg}
\label{sec:extent}

\begin{figure}
\centering \includegraphics[width=\columnwidth,angle=0]{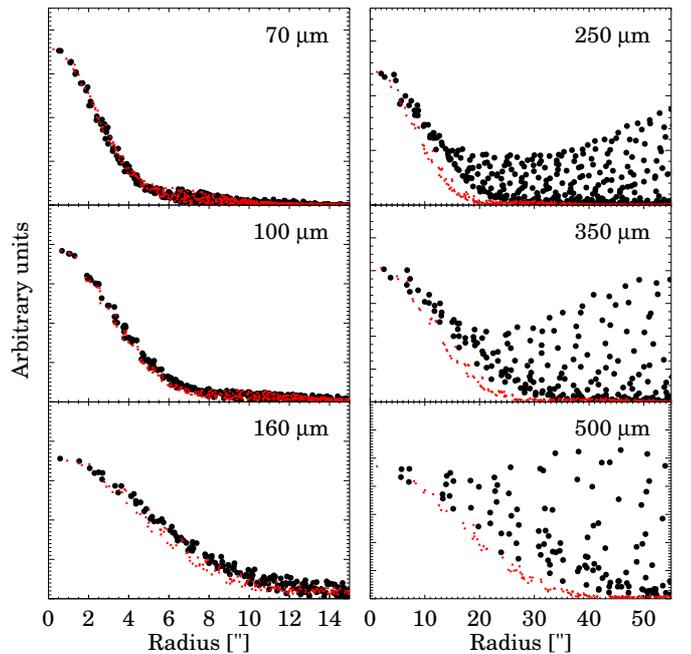}
\caption{Pixel values as a function of distance from the source
  centroid for the background-subtracted PACS images taken on 2011
  November 29 and for the SPIRE images taken on 2012 January 3 (black
  dots). Red dots were obtained in a similar way, but using
  observations of $\alpha$\,Boo for PACS and the official beam
  profiles for SPIRE.}
\label{fig:extent}
\end{figure}

In the following, we will determine whether the circumstellar matter
around V2492\,Cyg is resolved at any Herschel wavelength. In
Fig.~\ref{fig:extent} we plotted the pixel values of the
background-subtracted PACS images from 2011 November 29 as a function
of distance from the source centroid. The points can be well fitted
using Gaussians with full widths at half maximum (FWHMs) of
5$\farcs$2$\pm$0$\farcs$4, 6$\farcs$9$\pm$0$\farcs$5, and
11$\farcs$7$\pm$1$\farcs$4 at 70, 100, and 160$\,\mu$m,
respectively. As a point spread function (PSF) standard, we reduced
observations of $\alpha$\,Boo the same way as we reduced our
V2492\,Cyg observations and overplotted the obtained pixel values on
the same graphs. Figure~\ref{fig:extent} shows that while the source
is unresolved at 70$\,\mu$m, it is marginally resolved at longer
wavelengths. We calculated by how much the PSF needs to be broadened
to reproduce the observed profiles. From this, we could derive an
upper limit for the deconvolved size of $<$1$\farcs$1 ($<$580\,AU) at
70$\,\mu$m and deconvolved sizes of 2$\farcs$1 (1200\,AU) at
100$\,\mu$m and 4$\farcs$9 (2700\,AU) at 160$\,\mu$m. The PACS images
taken during the three other epochs give identical results. We also
fitted 2D Gaussians to the PACS images. The obtained major axes are on
average 5$\farcs$8, 7$\farcs$6, and 13$\farcs$0, confirming our
earlier findings.

Figure~\ref{fig:extent} also shows the pixel values measured on our
SPIRE images from 2012 January 3. Because we did not attempt to
perform any background subtraction, several points deviate from a
Gaussian profile due to large-scale extended emission. Here we used
the official SPIRE beam profiles provided for a 1$''$ resolution at
the Herschel Science Centre calibration ftp site, resampled them to
the same pixel size as our V2492\,Cyg observations, and overplotted
them on the same figure. In order to estimate the size of our source,
we again broadened the PSFs until we could reproduce the lower
envelope of the data points measured for V2492\,Cyg. The resulting
deconvolved sizes are 15$\farcs$3 (8400\,AU) at 250$\,\mu$m and
20$\farcs$0 (11\,000\,AU) at 350$\,\mu$m. The contrast between the
source and the surrounding extended emission is so low at 500$\,\mu$m
that no reasonable size estimate could be done.

\subsection{The broader environment of V2492\,Cyg}
\label{sec:broader}

\begin{figure*}
\centering \includegraphics[width=\textwidth,angle=0]{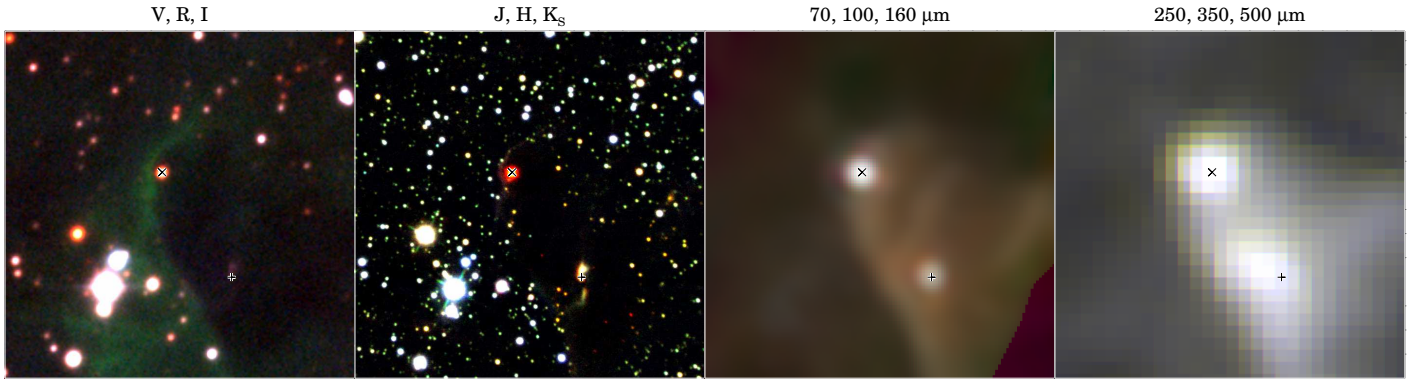}
\caption{False-color composites depicting the environment of
  V2492\,Cyg. Each image shows the same 3$'\,{\times}\,$3$'$ area. The
  two leftmost images are linearly scaled, the two rightmost ones are
  logarithmically scaled. The $V$, $R$, $I$ composite was constructed
  by shifting and co-adding several exposures taken with the Schmidt
  telescope at Konkoly Observatory, obtained between 2011 November
  1-30. The $J$, $H$, $K_{\rm S}$ composite was obtained with the WHT
  on 2011 July 22. The 70, 100, 160$\,\mu$m composite uses all
  \emph{Herschel}/PACS images taken between 2011 October 29 and 2012
  January 11. The 250, 350, 500$\,\mu$m composite is made of
  \emph{Herschel}/SPIRE images taken on 2012 January 3. Gray $\times$
  symbols mark the position of V2492\,Cyg ($\alpha_{2000}$=20$^{\rm
    h}$51$^{\rm m}$26$\fs$19,
  $\delta_{2000}$=+44$^{\circ}$05$'$23$\farcs$6). Gray $+$ symbols
  mark the position of HH\,570\,IRS ($\alpha_{2000}$=20$^{\rm
    h}$51$^{\rm m}$22$\fs$85,
  $\delta_{2000}$=44$^{\circ}$04$'$29$\farcs$51).}
\label{fig:panels}
\end{figure*}

V2492\,Cyg sits at the tip of a dark cloud that blocks out almost
every background star at optical wavelengths
(Fig.~\ref{fig:panels}). Because this cloud is illuminated from the
east and emits at H$\alpha$ \citep{bally2003}, the rim of the cloud is
visible in our $R$ band images, but not in the $V$ or $I$
bands. Interestingly, the rim is also faintly visible in the $J$, $H$,
and $K_{\rm S}$ bands. Due to the lower extinction at infrared
wavelengths compared to the optical, background stars appear behind
the cloud in the $JHK_{\rm S}$ images, making it possible to estimate
interstellar extinction in the vicinity of V2492\,Cyg. Using the 2MASS
magnitudes of the brightest stars as calibrators, we obtained
photometry for every point source brighter than $J$=20.5\,mag in the
UKIDSS $JHK_{\rm S}$ images \citep{kospal2011} in an area of
3$'\,{\times}\,$3$'$ centered on V2492\,Cyg. Then, we dereddened the
sources using the reddening law from \citet{cardelli1989} so that they
fall on the main sequence. We found that stars behind the dark cloud
typically have reddenings in the $A_V$=10--20\,mag range, while
outside the cloud, stars are typically reddened by only
$A_V$=1--10\,mag.

Moving to longer wavelengths, the cloud that is dark at optical and
infrared wavelengths starts emitting its own thermal radiation (see
the Herschel images in Fig.~\ref{fig:panels}). Using the convolution
kernels of \citet{gordon2008}, we convolved each Herschel image to the
resolution of the SPIRE 500$\,\mu$m images and then fitted a modified
blackbody to the observed SED for each map pixel. We used a fixed
powerlaw index of the dust opacity coefficient $\beta$ = 2 (assuming a
dust opacity law of $\kappa\,{\propto}\,\lambda^{-\beta}$),
appropriate for the diffuse interstellar medium \citep{draine2006}. We
found that the temperature across the cloud is rather homogeneous
(between 15 and 18\,K), coldest in the central parts and increasing
towards the edge of the cloud. This temperature is typical of dense
molecular clouds \citep[e.g.,][]{stahlerpalla}. Using our temperature
map and assuming optically thin emission, we calculated the optical
depth in the cloud and found values up to $\tau_{160}$=0.01. Taking
3000\,mag as a typical ratio between the extinction in the V-band and
the optical depth at 160$\,\mu$m \citep{kiss2006}, this optical depth
translates to $A_V$=30\,mag in the densest parts of the cloud, which
is roughly consistent with what we obtained from dereddening
background stars.

\citet{bally2003} reported the detection of several Herbig-Haro
objects in the Pelican Nebula. HH\,569 is a bright bow shock
supposedly originating from V2492\,Cyg. This object must only emit in
shock lines like [SII] because it is not visible either in our broad
band $VRIJHK_{\rm S}$ images or in our Herschel images. According to
\citet{bally2003}, HH\,570\,S, HH\,570, and HH\,570\,N form a single,
highly collimated flow, with no infrared point sources cataloged along
it. While we do not detect the southern and northern parts, the
central part, HH\,570, is well visible in all our images. In the $J$,
$H$, and $K_{\rm S}$ bands, the object seems to consist of a central
point source, a slightly curved conical nebula extending to about
5$''$ to the north, and a narrow southern nebula similar in size but
separated by a dark area from the central point source. We note that
this object, henceforth HH\,570\,IRS ($\alpha_{2000}$=20$^{\rm
  h}$51$^{\rm m}$22$\fs$85,
$\delta_{2000}$=44$^{\circ}$04$'$29$\farcs$51), has not been mentioned
in the literature before. The morphology suggests that the source is
likely a deeply embedded young stellar object surrounded by a bipolar
reflection nebula. HH\,570\,IRS is also well visible as a point source
in our Herschel/PACS images. This source may very well be the driving
source of HH\,570\,S and HH\,570\,N.

Interestingly, another faint point source also appears in our PACS
70$\,\mu$m images in the vicinity of V2492\,Cyg. This source is
located at a distance of 9$\farcs$9 to the west of V2492\,Cyg (P.A.~of
243$^{\circ}$) and is only visible at 70$\,\mu$m. The brightness
contrast between them is about 30. The faint source is not detected at
any other wavelengths. It may be an infrared companion to V2492\,Cyg
or an unrelated background object.

\section{Discussion}
\label{sec:dis}

In the following, we discuss what we can learn about the origin of
light variations in V2492\,Cyg. In Sect.~\ref{sec:light} we showed
that the light curves in Fig.~\ref{fig:light} have practically
identical shapes, although with different variability amplitudes, at
all wavelengths between 0.55 and 4.5$\,\mu$m. This result and the fact
that no time delays can be observed between the different curves
suggest a single physical mechanism behind the variability. One
possibility would be that the luminosity of the outbursting central
source changes in time due to variable accretion, and this varying
illumination causes changes in the thermal emission of the disk. This
explanation, however, is unlikely because in Sect.~\ref{sec:sed} we
concluded, partly on the basis of our Herschel monitoring program,
that the intrinsic brightness of the central source has been nearly
constant since the peak brightness in 2010 August. Moreover, a varying
central illumination would result in more wavelength-independent
brightness changes than observed. In addition, it might result in time
delays at longer wavelengths with respect to the changes in the
optical light curve (see, e.g., the case of OO\,Ser in
\citealt{kospal2007} and V1647\,Ori in \citealt{muzerolle2005,
  mosoni2012}). Based on our observations, we think that the picture
of time-dependent obscuration of the central part of the system by an
orbiting dust cloud in a nearly edge-on disk is more likely, both
because the color changes follow the extinction curve of
\citet{cardelli1989} and because there is a lack of time delay between
the light curves. Concerning its physics, this scenario is the same as
the one used to explain the minima of UX\, Orionis-type stars (UXors),
isolated Herbig stars showing sporadic 1-3\,mag fadings lasting for
weeks \citep{grinin1991, eaton1994}. We will adopt this explanation
for the forthcoming discussions.

To test the plausibility of the orbiting dust cloud scenario, we will
first check if the orbital velocity of the cloud is consistent with
the assumed Keplerian velocity field. Following the method used by
\citet{vanboekel2010} for T\,Tau\,S, we will assume that the measured
4.5$\,\mu$m brightness comes from a 1500\,K blackbody. Considering
that V2492\,Cyg faded by $F_{\nu}=$0.17\,Jy at 4.5$\,\mu$m within
6.8\,days, between 2012 January 4 and 11, the minimum solid angle of
the region that was covered is $\Omega = F_{\nu} / B_{\nu}(1500\,\rm
K) \approx 3 \times 10^{-18}\,$sr. This corresponds to a diameter of
0.2\,AU if the emitting region is circular. The velocity required to
obscure such an area in 6.8 days is 55\,km/s. This is not an
unreasonable value, since the Keplerian velocity around a star of
0.4\,M$_{\odot}$ at a radius of 0.1\,AU would be 55\,km/s. Even if the
4.5$\,\mu$m emission originates from the disk region somewhat cooler
than the assumed 1500\,K due to a radially decreasing temperature,
and thus brightness, profile, it is still the inner disk that needs
to be obscured. This result strongly supports that the obscuring cloud
must be part of the inner disk/envelope within a few tenths of an AU,
and dusty structures in the outer disk passing through the
line of sight can be excluded. We can make a simple estimate for the
mass of the obscuring cloud by considering that it causes at most
$A_V\sim$20\,mag extinction (Fig.~\ref{fig:tcd}). Taking the relation
between optical extinction and hydrogen column density from
\citet{guver2009}, $A_V\sim$20\,mag corresponds to a column density of
0.07\,g\,cm$^{-2}$ of gas and dust. Assuming a simple slab geometry
with a radius of 0.1\,AU, the total cloud mass is in the order of
10$^{-10}$\,M$_{\odot}$. Comparing this value with the total estimated
circumstellar mass of 0.06\,M$_{\odot}$ \citep{hillenbrand2012}, this
can be considered as a small mass inhomogeneity.

It is noteworthy that the optical variability amplitude of V2492\,Cyg
is significantly higher than typical minima depths of UXors. UXors
usually show only 1-3 mag fadings and exhibit a characteristic blueing
in the faintest state due to the increasing contribution of scattered
light from the circumstellar matter to the total brightness of the
system. This scattered-light component also naturally limits the
observed amplitudes of the brightness variations of UXors
\citep{grinin1988, natta2000}. That V2492\,Cyg exhibits much deeper
minima may be related to the fact that the obscuring cloud in this
system is probably larger. It covers not only the central star but
also the inner part of the circumstellar matter, thus eclipsing any
scattered light from the system. In this respect, V2492\,Cyg can be
regarded as a young and embedded analog of UXors.

We speculate about two possible scenarios concerning the origin of the
obscuring structure. One is a pre-existing, long-lived orbiting dust
structure that moves in/out of the line of sight. The other is a
transient appearance/disappearance of dust clouds in the system, due
either to dust condensation/evaporation driven by the changing
accretion heating or released from an accretion disk surface by
turbulence \citep{turner2010}. Due to its large opacity, the cloud
casts a shadow onto the outer parts of the circumstellar matter,
decreasing the far-infrared thermal emission of the shadowed area. In
the case of the permanent orbiting cloud, this shadow would not change
the total integrated far-infrared flux. However, in the transient
case, depending on the solid angle as seen from the star, detectable
far-infrared flux changes may occur (as was seen in the case of
PV\,Cep, see Fig.~3 of \citealt{kun2011}). Since part of the
far-infrared radiation comes from an optically thin medium (envelope,
disk surface), the flux changes should be synchronous with the
optical/near-infrared flux changes. In our 70$\,\mu$m \emph{Herschel}
monitoring (Sect.~\ref{sec:herschelmonitoring}), we observed no
detectable flux changes at 70$\,\mu$m or 160$\,\mu$m. Although this
result does not completely exclude the transient scenario (if the
solid angle of the cloud is small enough or the relative contribution
of optically thin emission to the total far-infrared flux is low), it
favors the long-lived dust structure. A consequence of the latter
scenario is some periodicity in the light variations, possibly
corresponding to the deep minima in Fig.~\ref{fig:light}. The
small-scale variability seen in our light curves might correspond to
smaller inhomogeneities within the obscuring cloud.

Although changing extinction is very likely the dominant factor in the
light variations of V2492\,Cyg, our observations indicate that there
are additional effects, too. Figure~\ref{fig:scatter} shows that while
the $VRI$ and 3.6-4.5$\,\mu$m magnitude changes are consistent with
the reddening, the $JHK_{\rm S}$ amplitudes are higher. In order to
explain this, we may consider a scenario where there is scattered
light from the directly illuminated side of the obscuring dust
cloud. When the cloud passes out of the line of sight and moves behind
the central source, the source becomes unobscured on one hand and on
the other, an additional scattered light component appears. Another
effect, expected to be most important at the same orbital phase, may
be extra near-infrared thermal radiation from the orbiting dust
cloud. These effects act in the same direction and could qualitatively
explain the larger $JHK_{\rm S}$ variability amplitudes seen in
Fig.~\ref{fig:scatter}.

We can check whether the observed spectral changes are consistent with
the picture described above. The most conspicuous change in the
near-infrared spectra was the disappearence of the H\,I lines during
the faint phase of the system in 2011 July-August, while the H$_2$
lines did not disappear, but became slightly stronger. Considering
that the atomic hydrogen lines originate from a hotter, denser
environment probably closer to the central star than where the
molecular hydrogen lines are emitted, it is possible that the H\,I
lines were simply obscured, probably extincted by the same cloud that
affected the continuum and broad-band photometry. Alternatively, the
lack of the Br$\,\gamma$ line could be interpreted as the cease of
accretion; however, the SED clearly shows that an outburst component,
i.e., the release of accretion energy, was still present in the faint
state. This apparent contradiction can be resolved if the H\,I lines
are related to the stellar wind rather than to the accretion, and
their disappearance signals the weakening of the wind. Indeed,
\citet{covey2011} and \citet{aspin2011} saw signatures of mass loss in
their spectra.

Finally, an interesting question is whether the existence of the
orbiting dust cloud is related to the outburst or whether eclipses
could occur during the quiescent phases as well. Our results, mainly
based on the \emph{Herschel} monitoring, suggest that the obscuring
cloud is not a short-lived temporary structure. Checking the
historical brightness values of V2492\,Cyg, \citet{hillenbrand2012}
concluded that in the past the source also exhibited light variations
in quiescence. Thus, the dust cloud might be present in the inner part
of the system for a longer time. The origin of such an asymmetric
structure also needs explanation because in an equilibrium disk,
density inhomogeneities would be quickly ironed out. In a similar
situation, \citet{muzerolle2009} argued that the orbiting warp
observed around the low-mass star LRLL\,31 might be maintained by a
closeby companion/planet. The structure of the obscuring dust cloud in
the V2492\,Cyg system, nevertheless, might have been changing during
the outburst, because the increased stellar wind and stellar radiation
could remove/evaporate dust particles that might move
back/recondensate in the lower intensity periods. This effect may
explain why the eclipses are not strictly periodic and why both the
time width and the amplitude of the fadings are variable.

\section{Summary and Conclusions}
\label{sec:con}

Using ground-based instrumentation as well as the \emph{Spitzer} and
\emph{Herschel} space telescopes, we obtained new multi-epoch
observations between 0.55$\,\mu$m and 160$\,\mu$m of the young
eruptive star V2492\,Cyg. We describe the flux evolution in the
optical--infrared domain between the peak brightness in 2010 August
and 2012 December, and we analyze the circumstellar and
interstellar environment of the star. Our main results and conclusions
are the following:

\begin{itemize}
\item V2492\,Cyg is a young stellar object located at the tip of a
  small dark cloud, which was mapped by our \emph{Herschel}
  observations. The densest parts of the cloud have an extinction of
  up to 30\,mag and a temperature of 15\,K. This cloud contains
  another young stellar object, the newly discovered HH\,570\,IRS.
\item The light curves of V2492\,Cyg show large amplitude variability
  including two very deep minima. Our analysis revealed that a single
  physical mechanism is responsible for the brightness changes, and
  the color variations suggest with high accuracy that the most likely
  explanation is changing extinction along the line of sight. In
  this respect, V2492\,Cyg can be regarded as a young, embedded
  analog of UX\,Orionis-type variables.
\item We propose that the obscuring structure is a non-axisymmetric
  dust cloud orbiting in the inner disk. Its optical thickness is in
  the 13--20\,mag range. Our order-of-magnitude estimates suggest that
  its mass is about 10$^{-10}$\,M$_{\odot}$, and it is orbiting at a
  few tenths of an AU from the star. The eclipses indicate that the
  geometry of the system is closer to edge-on than to pole-on. The
  disappearance of the H\,I lines during the faintest state also
  supports the obscuration scenario.
\item We found that the source exhibits higher amplitude variations in
  the near-infrared than expected from the optical amplitudes assuming
  a standard extinction curve. We speculate that scattered light as
  well as hot dust emission from the directly illuminated side of the
  orbiting cloud is responsible for this.
\item In order to decide whether the orbiting dust cloud is a
  temporary feature or a long-lived structure, we monitored the
  far-infrared flux with Herschel. The results are more consistent
  with a relatively long-lived feature.
\item The brightening of the source before 2010 August could not be
  explained by extinction changes. Our data show a flux increase in
  the whole 0.55--70$\,\mu$m wavelength range and suggest that the
  outburst component is not a single temperature blackbody. This
  component seems to be mostly invariant since the epoch of the peak
  brightness, suggesting a long-term ($>$2\,yr) eruption.
\end{itemize}

\begin{acknowledgements}

We thank the referee, Dr. Vladimir Grinin, for his valuable
comments, in particular those emphasizing the physical similarity
between V2492\,Cyg and the UX\,Orionis-type stars.
We acknowledge with thanks the variable star observations from the
AAVSO International Database contributed by observers worldwide and
used in this research, including recent monitoring by James Roe
obtained throu the Lowell Amateur Research Initiative.
This work was partly supported by the grants OTKA-101393, OTKA-104607,
OTKA K-81373, and OTKA K-81966 of the Hungarian Scientific Research
Fund, by the Lend\"ulet-2009 Young Researchers' Program of the
Hungarian Academy of Sciences, by the HUMAN MB08C 81013 grant of
the MAG Zrt, by the J\'anos Bolyai Research Scholarship of the Hungarian
Academy of Sciences, and by the grant LP2012-31/2012.
The William Herschel Telescope and its service program are operated
on the island of La Palma by the Isaac Newton Group in the Spanish
Observatorio del Roque de los Muchachos of the Instituto de
Astrof\'\i{}sica de Canarias.
This work is based in part on observations made with the Telescopio
Carlos S\'anchez operated on the island of Tenerife by the Instituto
de Astrof\'\i{}sica de Canarias in the Observatorio del Teide. The
authors wish to thank the telescope manager A.~Oscoz, the support
astronomers and telescope operators for their help during the
observations, as well as the service mode observers.
PACS has been developed by a consortium of institutes led by MPE
(Germany) and including UVIE (Austria); KUL, CSL, IMEC (Belgium); CEA,
OAMP (France); MPIA (Germany); IFSI, OAP/AOT, OAA/CAISMI, LENS, SISSA
(Italy); IAC (Spain). This development has been supported by the
funding agencies BMVIT (Austria), ESA-PRODEX (Belgium), CEA/CNES
(France), DLR (Germany), ASI (Italy), and CICT/MCT (Spain).
SPIRE has been developed by a consortium of institutes led by Cardiff
Univ. (UK) and including Univ. Lethbridge (Canada); NAOC (China); CEA,
LAM (France); IFSI, Univ. Padua (Italy); IAC (Spain); Stockholm
Observatory (Sweden); Imperial College London, RAL, UCL-MSSL, UKATC,
Univ. Sussex (UK); and Caltech, JPL, NHSC, Univ. Colorado (USA). This
development has been supported by national funding agencies: CSA
(Canada); NAOC (China); CEA, CNES, CNRS (France); ASI (Italy); MCINN
(Spain); SNSB (Sweden); STFC (UK); and NASA (USA).

\end{acknowledgements}

\bibliographystyle{aa}
\bibliography{paper}{}


\Online
\vspace*{70mm}
\hspace*{65mm}
\mbox{\LARGE{Online Material}}

\longtab{1}{
\begin{longtable}{lcccccccc}
\caption{\label{tab:phot} Optical and near-infrared photometry in magnitudes for
  V2492\,Cyg.}\\
\hline\hline
Date & JD$\,{-}\,$2\,400\,000 & $V$ & $R$    & $I$     & $J$      & $H$      & $K_{\rm S}$    & Telescope\\
\hline
\endfirsthead
\caption{continued.}\\
\hline\hline
Date & JD$\,{-}\,$2\,400\,000 & $V$ & $R$    & $I$     & $J$      & $H$      & $K_{\rm S}$    & Telescope\\
\hline
\endhead
\hline
\endfoot
2011-Mar-24 & 55\,644.61 & 19.16(6) & 17.58(4) & 15.71(4) &          &          &          & Schmidt \\
2011-Mar-25 & 55\,645.59 & 19.57(15)& 17.59(4) & 16.07(4) &          &          &          & RCC     \\
2011-Mar-30 & 55\,650.57 & 18.93(5) & 17.08(4) & 15.77(5) &          &          &          & RCC     \\
2011-Mar-31 & 55\,651.55 & 19.18(5) & 17.20(4) & 15.85(4) &          &          &          & RCC     \\
2011-Apr-10 & 55\,661.54 & 18.97(5) & 17.11(4) & 15.31(4) &          &          &          & Schmidt \\
2011-Apr-11 & 55\,662.59 & 18.46(9) & 16.59(4) & 14.90(4) &          &          &          & Schmidt \\
2011-Apr-23 & 55\,674.53 & 18.10(4) & 16.04(4) & 14.74(4) &          &          &          & RCC     \\
2011-Apr-27 & 55\,678.67 & 17.71(4) & 16.01(4) &          &          &          &          & IAC-80  \\
2011-Apr-27 & 55\,678.73 &          &          &          & 11.89(1) & 10.31(1) & 8.73(1)  & TCS     \\
2011-Jun-05 & 55\,717.53 &          & 18.36(9) & 17.70(4) &          &          &          & RCC     \\
2011-Jun-19 & 55\,732.39 & $>$18.6  & $>$19.8  & $>$18.9  &          &          &          & Schmidt \\
2011-Jun-20 & 55\,733.40 & $>$19.7  & $>$19.8  & $>$19.4  &          &          &          & Schmidt \\
2011-Jun-26 & 55\,739.50 &          &          & 21.60(40)&          &          &          & RCC     \\
2011-Jul-21 & 55\,763.65 &          &          &          & 17.26(3) & 13.56(1) & 10.59(1) & TCS     \\
2011-Jul-22 & 55\,764.69 &          &          &          & 16.66(7) & 13.39(8) & 10.51(7) & WHT     \\
2011-Jul-25 & 55\,768.46 &          &          & 20.93(40)&          &          &          & Schmidt \\
2011-Aug-04 & 55\,777.56 &          &          &          & 15.41(1) & 12.51(1) & 10.01(1) & TCS     \\
2011-Aug-05 & 55\,779.37 &          &          & 20.08(20)&          &          &          & Schmidt \\
2011-Aug-11 & 55\,785.40 &          &          & 18.78(30)&          &          &          & Schmidt \\
2011-Aug-12 & 55\,786.38 &          &          & 19.24(10)&          &          &          & Schmidt \\
2011-Aug-14 & 55\,788.36 &          &          & 19.42(30)&          &          &          & Schmidt \\
2011-Aug-15 & 55\,789.46 &          &          &          & 14.72(2) & 12.13(1) & 9.85(1)  & TCS     \\
2011-Aug-17 & 55\,791.35 &          &          & 18.60(30)&          &          &          & Schmidt \\
2011-Aug-18 & 55\,792.57 &          &          & 18.66(16)&          &          &          & Schmidt \\
2011-Aug-20 & 55\,794.51 &          &          & 18.16(5) &          &          &          & Schmidt \\
2011-Aug-23 & 55\,797.58 &          &          & 17.27(4) &          &          &          & Schmidt \\
2011-Aug-25 & 55\,799.35 &          & 19.05(40)& 16.52(10)&          &          &          & RCC     \\
2011-Aug-29 & 55\,803.35 & 21.55(40)& 18.14(20)& 16.37(10)&          &          &          & RCC     \\
2011-Aug-30 & 55\,804.34 & 21.88(40)& 18.12(20)& 16.48(10)&          &          &          & RCC     \\
2011-Sep-03 & 55\,807.53 &          &          & 16.01(40)&          &          &          & Schmidt \\
2011-Sep-06 & 55\,811.34 & 21.91(15)& 18.51(4) & 15.82(4) &          &          &          & Schmidt \\
2011-Sep-09 & 55\,813.60 &          &          &          & 12.40(1) & 10.60(2) & 8.83(1)  & TCS     \\
2011-Sep-09 & 55\,814.31 & 20.53(30)& 16.95(40)& 15.40(5) &          &          &          & Schmidt \\
2011-Sep-09 & 55\,814.45 &          &          &          & 12.32(3) & 10.52(1) & 8.79(1)  & TCS     \\
2011-Sep-10 & 55\,815.30 & 20.55(40)& 18.90(30)& 15.16(4) &          &          &          & RCC     \\
2011-Sep-11 & 55\,816.46 &          &          &          & 12.38(1) & 10.59(1) & 8.89(1)  & TCS     \\
2011-Sep-12 & 55\,817.47 &          &          &          & 12.39(1) & 10.54(1) & 8.81(1)  & TCS     \\
2011-Sep-13 & 55\,818.49 &          &          &          & 12.44(1) & 10.58(1) & 8.86(2)  & TCS     \\
2011-Sep-15 & 55\,820.31 & 20.15(25)& 18.48(25)& 15.95(6) &          &          &          & Schmidt \\
2011-Sep-16 & 55\,821.29 & 21.78(25)& 18.08(30)& 16.07(5) &          &          &          & Schmidt \\
2011-Sep-17 & 55\,822.30 & 22.48(17)& 18.30(25)& 16.03(4) &          &          &          & Schmidt \\
2011-Sep-21 & 55\,826.28 & 20.49(21)& 17.96(30)& 15.74(4) &          &          &          & Schmidt \\
2011-Sep-22 & 55\,827.35 & 20.26(10)& 17.78(30)& 15.57(4) &          &          &          & Schmidt \\
2011-Sep-23 & 55\,828.30 &          & 17.52(20)& 15.33(4) &          &          &          & Schmidt \\
2011-Sep-24 & 55\,829.30 & 19.26(10)& 16.69(4) & 15.10(4) &          &          &          & RCC     \\
2011-Sep-25 & 55\,830.29 & 20.09(18)& 17.49(14)& 15.25(4) &          &          &          & Schmidt \\
2011-Sep-25 & 55\,830.33 & 19.55(17)& 16.85(25)& 15.24(5) &          &          &          & RCC     \\
2011-Sep-26 & 55\,831.27 & 19.75(12)& 16.24(5) & 15.08(4) &          &          &          & Schmidt \\
2011-Sep-26 & 55\,831.29 & 19.53(8) & 16.59(17)& 15.12(4) &          &          &          & RCC     \\
2011-Sep-28 & 55\,833.25 & 19.17(5) & 16.84(4) & 14.83(4) &          &          &          & Schmidt \\
2011-Oct-07 & 55\,842.45 &          &          &          & 12.76(2) & 10.86(1) & 9.03(2)  & TCS     \\
2011-Oct-08 & 55\,843.45 &          &          &          & 12.66(1) & 10.77(1) & 8.97(1)  & TCS     \\
2011-Oct-11 & 55\,846.49 &          &          &          & 12.48(1) & 10.72(1) & 8.95(1)  & TCS     \\
2011-Oct-12 & 55\,846.50 &          &          &          & 12.47(1) & 10.72(1) & 8.95(1)  & TCS     \\
2011-Oct-26 & 55\,861.32 & 18.93(20)& 16.86(10)& 14.92(9) &          &          &          & Schmidt \\
2011-Oct-27 & 55\,862.30 & 18.84(10)& 17.07(8) & 15.07(4) &          &          &          & Schmidt \\
2011-Oct-27 & 55\,862.44 &          &          &          & 12.25(3) & 10.36(1) & 8.61(1)  & TCS     \\
2011-Oct-28 & 55\,863.22 & 19.44(7) & 17.54(4) & 15.51(4) &          &          &          & Schmidt \\
2011-Oct-29 & 55\,864.40 &          &          &          & 12.66(1) & 10.63(1) & 8.76(1)  & TCS     \\
2011-Oct-29 & 55\,864.41 & 19.73(11)& 17.49(5) & 15.58(4) &          &          &          & Schmidt \\
2011-Oct-30 & 55\,865.38 & 19.50(11)& 17.49(5) & 15.61(4) &          &          &          & Schmidt \\
2011-Oct-31 & 55\,866.37 &          &          &          & 12.85(1) & 10.72(1) & 8.84(1)  & TCS     \\
2011-Oct-31 & 55\,866.42 & 19.56(10)& 17.68(13)& 15.77(4) &          &          &          & Schmidt \\
2011-Nov-01 & 55\,867.21 & 19.62(13)& 17.57(4) & 15.63(4) &          &          &          & Schmidt \\
2011-Nov-02 & 55\,868.20 & 19.27(8) & 17.35(4) & 15.47(4) &          &          &          & Schmidt \\
2011-Nov-03 & 55\,869.34 & 19.07(10)& 17.07(6) & 15.15(4) &          &          &          & Schmidt \\
2011-Nov-04 & 55\,870.37 &          &          &          & 12.21(1) & 10.31(1) & 8.56(1)  & TCS     \\
2011-Nov-05 & 55\,971.19 & 18.63(5) & 16.71(4) & 14.84(4) &          &          &          & Schmidt \\
2011-Nov-05 & 55\,871.35 &          &          &          & 12.08(2) & 10.26(1) & 8.54(1)  & TCS     \\
2011-Nov-06 & 55\,872.19 & 18.64(11)& 16.63(4) & 14.77(4) &          &          &          & Schmidt \\
2011-Nov-06 & 55\,872.35 &          &          &          & 12.05(1) & 10.30(1) & 8.65(1)  & TCS     \\
2011-Nov-08 & 55\,874.19 & 18.34(13)& 16.21(4) & 14.36(4) &          &          &          & Schmidt \\
2011-Nov-08 & 55\,874.36 &          &          &          & 11.67(1) & 10.03(1) & 8.45(1)  & TCS     \\
2011-Nov-09 & 55\,875.18 & 18.24(10)& 16.23(4) & 14.36(4) &          &          &          & Schmidt \\
2011-Nov-13 & 55\,879.22 & 17.46(4) & 15.26(4) & 13.99(4) &          &          &          & RCC     \\
2011-Nov-22 & 55\,888.20 & 17.99(4) & 16.20(4) & 14.43(4) &          &          &          & Schmidt \\
2011-Nov-23 & 55\,889.20 & 18.24(4) & 16.24(4) & 14.50(4) &          &          &          & Schmidt \\
2011-Nov-24 & 55\,890.20 & 18.08(4) & 16.22(4) & 14.50(4) &          &          &          & Schmidt \\
2011-Nov-25 & 55\,891.25 & 18.45(4) & 16.54(4) & 14.74(4) &          &          &          & Schmidt \\
2011-Nov-27 & 55\,893.30 & 19.15(4) & 17.22(4) & 15.41(4) &          &          &          & Schmidt \\
2011-Nov-28 & 55\,894.19 & 19.30(10)& 17.44(4) & 15.60(4) &          &          &          & Schmidt \\
2011-Nov-29 & 55\,895.20 & 19.67(15)& 17.38(5) & 15.61(4) &          &          &          & Schmidt \\
2011-Nov-29 & 55\,895.35 &          &          &          & 12.72(1) & 10.74(1) & 8.89(1)  & TCS     \\
2011-Nov-30 & 55\,896.20 & 19.57(15)& 17.58(4) & 15.71(4) &          &          &          & Schmidt \\
2011-Nov-30 & 55\,896.36 &          &          &          & 12.77(1) & 10.77(1) & 8.97(1)  & TCS     \\
2012-Jan-03 & 55\,930.21 &          &          & 20.26(18)&          &          &          & Schmidt \\
2012-Jan-04 & 55\,931.32 &          &          & 19.81(17)&          &          &          & IAC-80  \\
2012-Jan-07 & 55\,934.19 &          &          & $>$18.6  &          &          &          & RCC     \\
2012-Jan-09 & 55\,936.21 &          &          & $>$18.2  &          &          &          & RCC     \\
2012-Jan-11 & 55\,938.31 &          &          & $>$18.2  &          &          &          & IAC-80  \\
2012-Jan-11 & 55\,938.32 &          &          &          & 15.95(3) & 12.84(2) & 10.25(1) & TCS     \\
2012-Jan-23 & 55\,950.20 &          &          & $>$18.9  &          &          &          & RCC     \\
2012-Feb-11 & 55\,968.69 &          & $>$20.1  & $>$19.0  &          &          &          & Schmidt \\
2012-Apr-28 & 56\,045.58 &          &          & 17.48(5) &          &          &          & Schmidt \\
2012-May-19 & 56\,066.55 &          & 17.94(4) & 16.25(4) &          &          &          & RCC     \\
2012-May-19 & 56\,067.49 & 20.14(5) & 17.97(4) & 16.14(4) &          &          &          & RCC     \\
2012-May-23 & 56\,071.49 & 20.16(9) & 17.64(4) & 16.07(4) &          &          &          & RCC     \\
2012-Jun-01 & 56\,079.54 &          & 17.58(4) & 15.43(4) &          &          &          & Schmidt \\
2012-Jun-04 & 56\,082.54 & 19.08(19)& 17.22(5) & 15.05(4) &          &          &          & Schmidt \\
2012-Jun-06 & 56\,085.41 & 18.29(10)& 16.95(10)& 15.01(9) &          &          &          & Schmidt \\
2012-Jun-07 & 56\,086.44 &          &          & 15.16(10)&          &          &          & Schmidt \\
2012-Jun-11 & 56\,090.37 & 19.79(10)& 17.50(5) & 15.44(4) &          &          &          & Schmidt \\
2012-Jun-20 & 56\,098.51 & 20.50(17)& 17.80(7) & 15.80(3) &          &          &          & Schmidt \\
2012-Jun-23 & 56\,102.42 & 19.89(12)& 17.62(8) & 15.64(6) &          &          &          & Schmidt \\
2012-Jun-27 & 56\,105.54 &          & 17.36(4) & 15.44(4) &          &          &          & Schmidt \\
2012-Jun-27 & 56\,106.43 &          & 17.18(4) & 15.22(4) &          &          &          & Schmidt \\
2012-Jul-08 & 56\,117.34 & 18.09(15)& 16.04(4) & 14.30(4) &          &          &          & Schmidt \\
2012-Jul-10 & 56\,118.53 & 18.00(7) & 16.02(4) & 14.27(4) &          &          &          & Schmidt \\
2012-Jul-13 & 56\,122.44 & 18.14(6) & 16.18(4) & 14.38(4) &          &          &          & Schmidt \\
2012-Jul-15 & 56\,124.36 &          & 16.19(4) & 14.39(4) &          &          &          & Schmidt \\
2012-Jul-16 & 56\,125.43 &          & 16.35(4) & 14.53(4) &          &          &          & Schmidt \\
2012-Jul-17 & 56\,126.48 &          & 16.22(4) & 14.42(4) &          &          &          & Schmidt \\
2012-Jul-22 & 56\,131.34 & 17.48(12)& 15.60(4) & 13.93(4) &          &          &          & Schmidt \\
2012-Jul-23 & 56\,132.34 & 17.35(7) & 15.56(4) & 13.93(4) &          &          &          & Schmidt \\
2012-Jul-25 & 56\,134.45 & 17.46(4) & 15.50(4) & 14.11(4) &          &          &          & IAC-80  \\
2012-Jul-25 & 56\,134.61 &          &          &          & 11.62(1) & 10.03(1) & 8.47(1)  & TCS     \\
2012-Aug-09 & 56\,149.34 & 17.27(4) & 15.26(4) & 14.03(4) &          &          &          & RCC     \\
2012-Aug-19 & 56\,159.38 & 17.10(5) & 15.06(4) & 13.37(5) &          &          &          & RCC     \\
2012-Aug-20 & 56\,160.38 & 16.37(5) & 14.60(5) & 13.31(4) &          &          &          & IAC-80  \\
2012-Aug-21 & 56\,160.68 &          &          &          & 11.19(1) & 9.59(1)  & 8.09(1)  & TCS     \\
2012-Aug-23 & 56\,163.40 & 16.98(10)& 15.79(10)& 13.54(5) &          &          &          & RCC     \\
2012-Aug-24 & 56\,164.31 & 16.91(10)& 15.68(10)& 13.43(6) &          &          &          & RCC     \\
2012-Sep-16 & 56\,187.26 & 17.10(5) & 15.15(4) & 13.94(4) &          &          &          & RCC     \\
2012-Oct-09 & 56\,210.41 &          &          &          & 11.92(1) & 10.10(1) & 8.40(1)  & TCS     \\
2012-Oct-10 & 56\,211.46 &          &          &          & 12.01(1) & 10.20(1) & 8.45(1)  & TCS     \\
2012-Oct-11 & 56\,212.45 &          &          &          & 12.08(1) & 10.29(1) & 8.55(1)  & TCS     \\
2012-Oct-12 & 56\,213.41 &          &          &          & 12.05(1) & 10.28(1) & 8.57(1)  & TCS     \\

\end{longtable}}

\onltab{2}{
\begin{table*}
\caption{Mid-infrared and far-infrared photometry in Jy for
  V2492\,Cyg.}\label{tab:spitzerherschel}
\footnotesize
\begin{tabular}{l@{}c@{}cccccccccc}
\hline \hline
Date        & JD$\,{-}\,$2\,400\,000 & $F_{3.6}$       & $F_{4.5}$       & $F_{70}$       & $F_{100}$      & $F_{160}$      & $F_{250}$    & $F_{350}$   & $F_{500}$   & Telescope \\
\hline

2011-Sep-08 & 55\,813.01             & 0.869$\pm$0.038 & 1.562$\pm$0.070 &                &                &                &              &             &             & \emph{Spitzer}   \\
2011-Sep-24 & 55\,829.48             & 1.027$\pm$0.044 & 1.790$\pm$0.080 &                &                &                &              &             &             & \emph{Spitzer}   \\
2011-Oct-29 & 55\,864.37             &                 &                 & 14.80$\pm$0.60 &                & 18.02$\pm$1.60 &              &             &             & \emph{Herschel}  \\
2011-Nov-29 & 55\,895.32             & 1.022$\pm$0.044 & 1.823$\pm$0.080 &                &                &                &              &             &             & \emph{Spitzer}   \\
2011-Nov-29 & 55\,895.29             &                 &                 & 14.54$\pm$0.60 & 16.82$\pm$1.00 & 18.36$\pm$1.60 &              &             &             & \emph{Herschel}  \\
2012-Jan-03 & 55\,930.77             &                 &                 &                &                &                & 11.8$\pm$2.0 & 7.3$\pm$1.5 & 3.4$\pm$1.0 & \emph{Herschel}  \\
2012-Jan-04 & 55\,931.46             & 0.712$\pm$0.031 & 1.425$\pm$0.063 &                &                &                &              &             &             & \emph{Spitzer}   \\
2012-Jan-06 & 55\,933.39             &                 &                 & 14.20$\pm$0.60 &                & 17.51$\pm$1.60 &              &             &             & \emph{Herschel}  \\
2012-Jan-11 & 55\,938.25             & 0.607$\pm$0.027 & 1.258$\pm$0.056 &                &                &                &              &             &             & \emph{Spitzer}   \\
2012-Jan-11 & 55\,938.34             &                 &                 & 14.43$\pm$0.60 &                & 17.63$\pm$1.60 &              &             &             & \emph{Herschel}  \\
2012-Jul-25 & 56\,134.10             & 0.877$\pm$0.038 & 1.464$\pm$0.065 &                &                &                &              &             &             & \emph{Spitzer}   \\
2012-Aug-20 & 56\,160.45             & 1.309$\pm$0.056 & 2.057$\pm$0.091 &                &                &                &              &             &             & \emph{Spitzer}   \\
2012-Sep-16 & 56\,186.75             & 1.072$\pm$0.046 & 1.732$\pm$0.077 &                &                &                &              &             &             & \emph{Spitzer}   \\
2012-Oct-12 & 56\,212.14             & 1.007$\pm$0.043 & 1.649$\pm$0.073 &                &                &                &              &             &             & \emph{Spitzer}   \\
\hline
\end{tabular}
\tablefoot{All fluxes are color-corrected.}
\end{table*}
}

\end{document}